\documentclass[aps,twocolumn,prl,superscriptaddress,showpacs, longbibliography]{revtex4-1}
\usepackage{graphicx}
\usepackage{bm}
\usepackage{amsmath}
\usepackage{setspace}
\usepackage{amsfonts}
\usepackage{amssymb}
\usepackage{epstopdf}
\usepackage{natbib} 
\usepackage{xspace}
\usepackage{color}
\usepackage[colorlinks,bookmarks=false,citecolor=blue,linkcolor=blue,urlcolor=bl
ue]{hyperref}

\newcommand{\bacusio}{BaCuSi$_2$O$_6$\xspace}

\newcommand{\coordvec}[3]{$(#1\enspace #2\enspace #3)$\xspace}
\newcommand{\Qhdir}{\coordvec{Q_{h}}{0}{4}}
\newcommand{\QhQhdir}{\coordvec{Q_{h}}{Q_{h}}{4}}
\newcommand{\Qldir}{\coordvec{0}{0}{Q_{l}}}
\newcommand{\Qlldir}{\coordvec{1}{0}{Q_{l}}}

\newcommand{\Fig}[1]{Fig.~\ref{#1}\xspace}
\newcommand{\FigInd}[1]{(#1)\xspace}
\newcommand{\FigIndRange}[2]{(#1)-(#2)\xspace} 
\newcommand{\FigSub}[2]{Fig.~\ref{#1}(#2)}
\newcommand{\FigSubRange}[3]{Figs.~\ref{#1}(#2)-\ref{#1}(#3)} 

\newcommand{\A}{\textit{A}\xspace}
\newcommand{\B}{\textit{B}\xspace}
\newcommand{\C}{\textit{C}\xspace}
\newcommand{\BC}{\textit{B}+\textit{C}\xspace}

\newcommand{\JAJAval}{$J_{A}$ = 4.275(5)}
\newcommand{\JBJBval}{$J_{B}$ = 4.72(1)}
\newcommand{\JCJCval}{$J_{C}$ = 4.95(2) meV}
\newcommand{\JNNAJNNAval}{$J^{\prime}_{A}$ = $-$0.480(3)}
\newcommand{\JNNBJNNBval}{$J^{\prime}_{B}$ = $-$0.497(8)}
\newcommand{\JNNCJNNCval}{$J^{\prime}_{C}$ = $-$0.57(1) meV}
\newcommand{\JinterJinterval}{$J^{\prime\prime}$ = $-$0.04(1) meV} 

\newcommand{\FigOneDimCuts}{Fig.~3\xspace}

\newcommand{\refequ}[1]{Eq.~(\ref{#1})\xspace}

\newcommand{\lacuo}{La$_{2}$CuO$_{4}$}
\newcommand{\srcuo}{SrCuO$_{2}$}

\newcommand{\unprimedUC}{$\{\hat{a}, \hat{b}, \hat{c}\}$\xspace}
\newcommand{\primedUC}{$\{\hat{a}^{\,\prime}, \hat{b}^{\,\prime}, \hat{c}^{\,\prime}\}$\xspace}

\begin{document}

\title{Multiple Magnetic Bilayers and Unconventional Criticality without 
Frustration in \bacusio}

\author{S. Allenspach}
\affiliation{Neutrons and Muons Research Division, Paul Scherrer Institut, 
CH-5232 Villigen, Switzerland}
\affiliation{Department of Quantum Matter Physics, University of Geneva, 
CH-1211 Geneva, Switzerland}
\author{A. Biffin}
\affiliation{Laboratory for Neutron Scattering and Imaging, Paul Scherrer 
Institut, CH-5232 Villigen, Switzerland}
\author{U. Stuhr}
\affiliation{Laboratory for Neutron Scattering and Imaging, Paul Scherrer 
Institut, CH-5232 Villigen, Switzerland}
\author{G. S. Tucker}
\affiliation{Laboratory for Neutron Scattering and Imaging, Paul Scherrer 
Institut, CH-5232 Villigen, Switzerland}
\affiliation{Institute of Physics, \'{E}cole Polytechnique 
F\'{e}d\'{e}rale de Lausanne (EPFL), CH-1015 Lausanne, Switzerland}
\author{S. Ohira-Kawamura}
\affiliation{J-PARC Center, Japan Atomic Energy Agency, Tokai, Ibaraki 
319-1195, Japan}
\author{M. Kofu}
\affiliation{J-PARC Center, Japan Atomic Energy Agency, Tokai, Ibaraki 
319-1195, Japan}
\author{D. J. Voneshen}
\affiliation{ISIS Facility, Rutherford Appleton Laboratory, Chilton, Didcot 
OX11 0QX, United Kingdom}
\author{M. Boehm}
\affiliation{Institut Laue Langevin, 6 Rue Jules Horowitz BP156, 38024 
Grenoble Cedex 9, France}
\author{B. Normand}
\affiliation{Neutrons and Muons Research Division, Paul Scherrer Institut, 
CH-5232 Villigen, Switzerland}
\author{N. Laflorencie}
\affiliation{Laboratoire de Physique Th\'{e}orique, CNRS and Universit\'{e} 
de Toulouse, 31062 Toulouse, France}
\author{F. Mila}
\affiliation{Institute of Physics, \'{E}cole Polytechnique F\'{e}d\'{e}rale 
de Lausanne (EPFL), CH-1015 Lausanne, Switzerland}
\author{Ch. R\"{u}egg}
\affiliation{Neutrons and Muons Research Division, Paul Scherrer Institut, 
CH-5232 Villigen, Switzerland}
\affiliation{Department of Quantum Matter Physics, University of Geneva, 
CH-1211 Geneva, Switzerland}

\begin{abstract}
The dimerized quantum magnet \bacusio was proposed as an example of 
``dimensional reduction'' arising near the magnetic-field-induced 
quantum critical point (QCP) due to perfect geometrical frustration of 
its inter-bilayer interactions. We demonstrate by high-resolution neutron 
spectroscopy experiments that the effective intra-bilayer interactions are 
ferromagnetic, thereby excluding frustration. We explain the apparent 
dimensional reduction by establishing the presence of three magnetically 
inequivalent bilayers, with ratios 3:2:1, whose differing interaction 
parameters create an extra field-temperature scaling regime near the QCP 
with a non-trivial but non-universal exponent. We demonstrate by detailed 
quantum Monte Carlo simulations that the magnetic interaction parameters 
we deduce can account for all the measured properties of \bacusio, opening 
the way to a quantitative understanding of non-universal scaling in any 
modulated layered system.
\end{abstract}

\maketitle

\begin{figure}[t] 
\includegraphics[width=0.75\columnwidth]{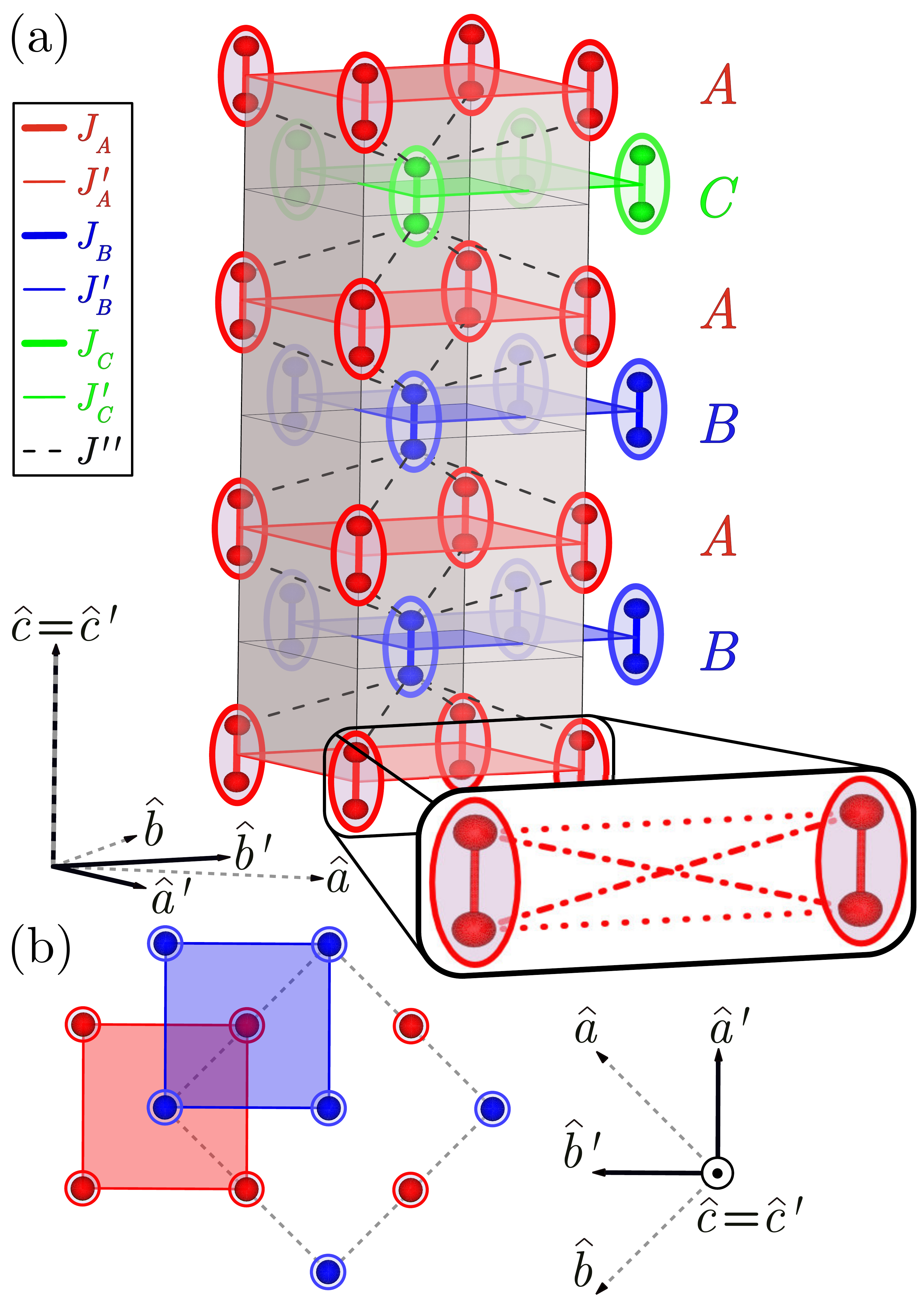}
\caption[]{\FigInd{a} Schematic representation of one unit cell of the 
minimal magnetic model for \bacusio; $\{J_\sigma, J_\sigma^\prime, J^{\prime\prime} \}$ 
are Heisenberg interactions. The three distinct bilayer types are labeled \A, 
\B, and \C. The effective interdimer interaction parameters within each 
bilayer ($J_\sigma^\prime$, edges of colored squares) result from four pairwise 
ionic interactions (inset). \FigInd{b} Top view of the $ab$ plane. The minimal 
unit cell, containing one dimer per bilayer \cite{Finger, Sasago}, has basis 
vectors $\{\hat{a}^{\,\prime}, \hat{b}^{\,\prime}, \hat{c}^{\,\prime}\}$; the 
crystallographic unit cell, used in previous scattering studies \cite{Samulon, 
MultipleMagnonModes, TwoTypesOfAdjacentLayers}, has basis $\{\hat{a}, \hat{b}, 
\hat{c}\}$ and contains two.} 
\label{fig:dimercell}
\end{figure}

A foundation stone of statistical physics is the theory of classical 
and quantum criticality \cite{ZinnJustin, Sachdev}, which states that all 
physical properties around a quantum phase transition (QPT) obey universal 
scaling laws dependent only on the dimension of space, $d$, and the dynamical 
exponent, $z$ (the ``dimension of time'' arising from the dispersion $\omega 
\propto k^z$ of low-energy excitations). The idea that perfectly frustrated 
competing interactions could lead to an effective reduction of $d$ has been 
both proposed \cite{Sebastian, BatistaFischer2007, SchmalianBatista2008} and 
contested \cite{Coleman, RoeschVojta2007_short, RoeschVojta2007_long} to 
explain the physics of \bacusio. This $S = 1/2$ material, known as Han 
purple from its use as a pigment in ancient China \cite{Finger, FitzHugh},
presents a three-dimensional (3D) stack of Cu$^{2+}$ bilayers 
[\FigSub{fig:dimercell}{a}] with dominant antiferromagnetic (AF) dimerization, 
significant intra-bilayer interactions, and a geometrically exact offset 
between adjacent bilayers, but was reported to show 2D scaling exponents 
around the field-induced QPT \cite{Sebastian}. The discovery of inequivalent 
bilayers in \bacusio \cite{MultipleMagnonModes, Kraemer2007} raised the 
question of whether frustration or structural modulation, or both, would be 
required to explain the apparent dimensional reduction \cite{Sebastian};
despite intensive investigation \cite{Kraemer2007, Kraemer2013, 
LaFlorencie2009, LaFlorencie2011, RoeschVojta2007_short, RoeschVojta2007_long, 
BatistaFischer2007, SchmalianBatista2008, TwoTypesOfAdjacentLayers}, this issue 
has yet to be resolved, with far-reaching implications for any layered material.

While field-driven QPTs from the ``quantum disordered'' dimerized state 
to the field-induced ordered state have remained a hot topic in quantum 
magnetism for multiple reasons (``Bose-Einstein condensation of magnons'') 
\cite{Zapf}, several recent developments make this the right time to revisit 
dimensional reduction in \bacusio. First, an {\it ab initio} analysis of the 
magnetic interactions has suggested that the effective intra-bilayer 
interactions are ferromagnetic (FM) \cite{Mazurenko}, which would preclude 
a frustration scenario. Second, a systematic structural determination 
\cite{TwoTypesOfAdjacentLayers} has confirmed at minimum two inequivalent 
and alternating bilayer units. Third, a new generation of time-of-flight 
(TOF) neutron spectrometers now allows magnetic excitations in materials 
such as \bacusio~to be characterized with unprecedentedly high resolution, 
and across multiple Brillouin zones.

In this Letter, we report the results of neutron spectroscopy experiments 
performed to determine the full magnetic Hamiltonian of \bacusio. We verify 
that the effective intra-bilayer interaction parameter is FM, establish the 
presence of three inequivalent bilayers with number ratios 3:2:1, and 
determine the very weak inter-bilayer interaction. We demonstrate by 
quantum Monte Carlo (QMC) simulations that our deduced interactions are 
completely consistent with all prior experimental data for the magnetization, 
phase diagram, layer triplet populations, and quantum critical behavior. Our 
conclusion that structural modulation creates an additional regime of 
unconventional effective scaling behavior will have broad applicability 
in the age of designer assembly of atomically thin magnetic materials.

The room-temperature crystal structure of \bacusio is tetragonal, but 
becomes weakly orthorhombic below 90 K \cite{Samulon, Sparta, Stern}. As 
\FigSub{fig:dimercell}{a} represents, the stacked ``bilayers'', square 
lattices of Cu$^{2+}$ spin dimers, have a relative shift of $\big( \frac{1}{2}$ 
$\frac{1}{2} \big)$ in the $ab$ plane [in the minimal unit cell of one 
bilayer \cite{Finger, Sasago}, represented in \FigSub{fig:dimercell}{b}].
The orthorhombic phase contains at least two structurally distinct bilayer 
types \cite{Samulon, TwoTypesOfAdjacentLayers}, presumably with different 
intra-bilayer interaction parameters, and with a very weak supercell 
structure showing incommensurate peaks \cite{Kraemer2007} near $Q_{k} = 
1/8$ \cite{Samulon}. Using the crystallographic unit cell \cite{Samulon, 
TwoTypesOfAdjacentLayers} [\FigSub{fig:dimercell}{b}] to interpret the 
measured dispersion relations \cite{MultipleMagnonModes} creates an ambiguity 
between FM and AF intra-bilayer interactions, as detailed in Sec.~S1 of the 
Supplemental Material (SM) \cite{SM}, \nocite{Mantid, Utsusemi, Horace, Haley, 
rnr, rmnrs, NeutronScatteringInCondensedMatterPhysics, ITC_VolumeC, 
La2CuO4_Paper1, La2CuO4_Paper2, SrCuO2_Paper1, SrCuO2_Paper2, Campostrini2000} 
which we resolve by working in the minimal unit cell. From the experimental 
map of the dynamical structure factor that we obtain over multiple Brillouin 
zones, we demonstrate that the minimal magnetic model is that shown in 
\FigSub{fig:dimercell}{a} and deduce the values of all the interaction 
parameters.

We have performed high-resolution inelastic neutron scattering (INS) 
measurements on the direct-geometry TOF spectrometers AMATERAS at the 
J-PARC neutron source \cite{AMATERAS} and LET at ISIS \cite{LET}. Respective 
measurement temperatures were 0.3 and 1.6 K. On both instruments, incident 
neutrons of energy $E_i = 9$ meV observe the full band width of the magnetic 
excitations (3--6 meV at zero applied field). The triple-axis spectrometers 
TASP and EIGER at the SINQ neutron source \cite{EIGER} were used for further 
investigation of selected $\vec{Q}$ directions, measuring at 1.6 K on both. 
All experiments used one \bacusio single crystal, of weight 1.01 g, which we 
discuss in Sec.~S2 of the SM \cite{SM}.

\begin{figure*}[t]
\begin{center}
\includegraphics[width=0.99\textwidth]{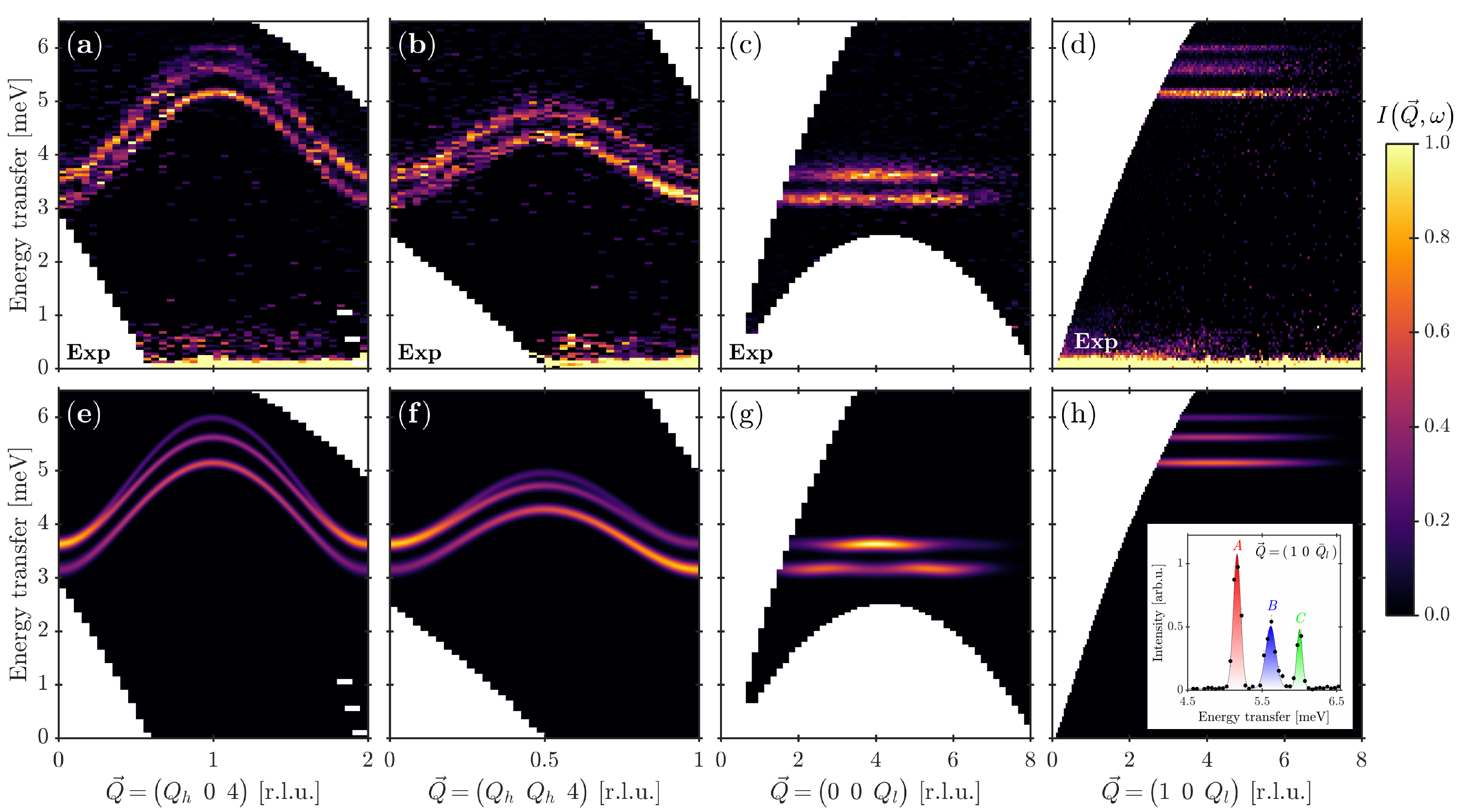}
\caption[]{\FigIndRange{a}{d} Intensity, $I({\vec Q}, \omega)$, measured 
on AMATERAS for selected high-symmetry directions. \FigIndRange{e}{h} 
Corresponding spectra calculated with the fitted interaction parameters. 
${\vec Q}$ is indexed in reciprocal lattice units (r.l.u.) of the crystallographic 
unit cell. Inset in \FigInd{h} shows the intensity, $I(\omega)$, obtained by 
integrating over the ${\vec Q}$ ranges [0.95, 1.05] in $Q_{h}$, [$-0.05$, 0.05] 
in $Q_k$, and [4, 8] in $Q_{l}$ (denoted by $\bar{Q}_{l}$).}
\label{fig:ExpAndSim}
\end{center}
\end{figure*}

\begin{figure}[t]
\begin{center}
\includegraphics[width=\columnwidth]{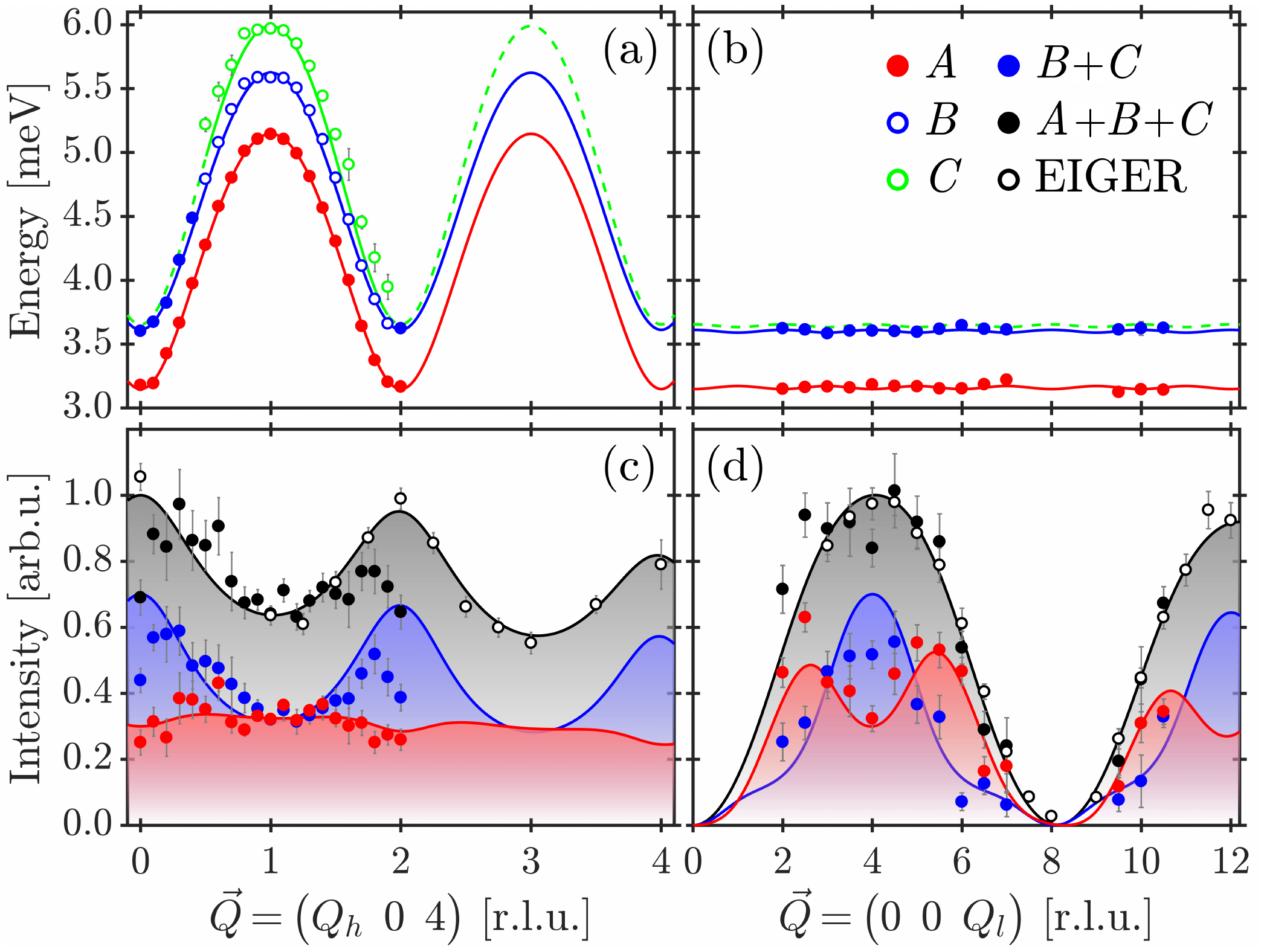}
\caption[]{Mode energies and intensities. All data other than the open 
circles in (c) and (d) are from AMATERAS. Lines show calculated 
results; in (a) and (b) the solid blue line represents \B, whereas in 
(c) and (d) it represents the contribution of \BC. The dashed green 
line denotes the energy of \C calculated in the range where it was not 
distinguishable in experiment.}
\label{fig:OneDimCuts}
\end{center}
\end{figure}

TOF intensity data for four high-symmetry directions in $\vec{Q}$ space are 
shown as functions of energy transfer in \FigSubRange{fig:ExpAndSim}{a}{d}. 
Figures \ref{fig:OneDimCuts}(a) and \ref{fig:OneDimCuts}(b) show the measured 
mode energies and Figs.~\ref{fig:OneDimCuts}(c) and \ref{fig:OneDimCuts}(d) 
their intensities, both obtained using Gaussian fits for selected 
$\vec{Q}$ points. The EIGER data integrate over a broad energy range to 
obtain high-statistics information for the combined mode intensities over an 
extended $|\vec{Q}|$ range. Details of data pre-processing and the Gaussian 
fitting procedure are presented in Sec.~S3 of the SM \cite{SM}.

Our results show clearly the presence of three distinct excitations 
in large regions of the Brillouin zone [Figs.~\ref{fig:ExpAndSim}(a), 
\ref{fig:ExpAndSim}(b), \ref{fig:ExpAndSim}(d), and \ref{fig:OneDimCuts}(a)]. 
These must result from three different types of bilayer, and so we label them 
\A, \B, and \C in ascending order of energy. This confirms the result of 
Ref.~\cite{MultipleMagnonModes}, but over a much wider $|\vec{Q}|$ range.
Where only two modes are visible because \B and \C are close in energy 
[Figs.~\ref{fig:ExpAndSim}(c) and \ref{fig:OneDimCuts}(b)], we label the 
effective composite mode \BC.

We draw attention to three qualitative features of our data, which all 
lie beyond the results of Ref.~\cite{MultipleMagnonModes}. (i) The minima 
of the strongly dispersive modes in Figs.~\ref{fig:ExpAndSim}(a), 
\ref{fig:ExpAndSim}(b), and \ref{fig:OneDimCuts}(a) give an unambiguous 
statement about the sign of the intra-bilayer interactions when working 
in the minimal unit cell. (ii) Although the inter-bilayer interaction is 
very weak, making the bands in \FigSub{fig:ExpAndSim}{c} almost flat, it can 
be determined from the variation of the intensity with $Q_l$. (iii) Where 
this band dispersion becomes $Q_l$ independent, in \FigSub{fig:ExpAndSim}{d}, 
the data can be used to establish the relative intensities of the three 
separate bilayer contributions. 

The magnetic excitations of the dimerized $S = 1/2$ system are 
``triplon'' quasiparticles. To model the triplon spectrum in \bacusio, we 
generalized the method of Refs.~\cite{Leuenberger1983, Leuenberger1984} as 
outlined in Sec.~S4 of the SM \cite{SM}. We assume that the only magnetic 
interactions are those of \FigSub{fig:dimercell}{a} and, given the weak 
anisotropies measured in \bacusio~\cite{Zvyagin}, that they have purely 
Heisenberg character. A quantitative fit of the complete mode energy and 
intensity data yields the intradimer interaction parameters \JAJAval, 
\JBJBval, and \JCJCval, intra-bilayer interactions \JNNAJNNAval, \JNNBJNNBval, 
and \JNNCJNNCval, and the inter-bilayer interaction \JinterJinterval, where 
$J > 0$ refers to AF interactions and $J < 0$ to FM. The quoted errors are 
statistical. These optimal values were used to calculate the spectra shown 
in \FigSubRange{fig:ExpAndSim}{e}{h} and as the lines in \Fig{fig:OneDimCuts}. 
As part of the fitting process, we used the high-$|{\vec Q}|$ EIGER data to 
deduce the anisotropic magnetic form factor (Sec.~S5 of the SM \cite{SM}), 
which confirms the concentration of spin density within the bilayers 
\cite{Mazurenko, Zvyagin}. We compare our interaction parameters with those 
deduced from previous INS studies \cite{MultipleMagnonModes, Sasago} in 
Sec.~S6 of the SM \cite{SM}.

Returning to the primary experimental observations, (i) the positions of 
the minima and maxima of the dispersive modes in \Qhdir and \QhQhdir are 
characteristic of FM effective intra-bilayer interactions. Because each 
$J_\sigma^\prime$ is the sum of four inter-ionic interactions, which are 
generally AF, their sign indicates that the diagonal interactions 
represented in the inset of \FigSub{fig:dimercell}{a} are dominant
\cite{Mazurenko}. For effective AF interactions, the positions of the 
maxima and minima would be exchanged; INS spectra calculated for intra- and 
inter-bilayer interactions of different signs are shown in Sec.~S6 of the 
SM \cite{SM}. We stress again that FM intra-bilayer interactions mean there 
is no inter-bilayer frustration in \bacusio.

(ii) While the very weak $J^{\prime\prime}$ results in almost flat modes along 
\Qldir [Figs.~\ref{fig:ExpAndSim}(c) and \ref{fig:OneDimCuts}(b)], the mode 
intensities are $Q_{l}$ dependent, displaying a double-peak structure in mode 
\A with maxima at $Q_{l}$ = 3 and 5 [Figs.~\ref{fig:ExpAndSim}(c) and 
\ref{fig:OneDimCuts}(d)]. This feature, which allows us to fit 
$J^{\prime\prime}$, could not be resolved at all in 
Ref.~\cite{MultipleMagnonModes} and was revealed only by using 
optimized chopper settings on both high-resolution TOF spectrometers. 
The fact that the double peak appears in \A is a direct consequence of a FM 
$J^{\prime\prime}$; an AF inter-bilayer interaction would cause it to appear in 
\BC (Sec.~S6 of the SM \cite{SM}).

(iii) Because the \Qlldir dispersion [\FigSub{fig:ExpAndSim}{d}] is 
$Q_l$ invariant, it is of particular value for a quantitative determination 
of the relative fractions of each bilayer type. The large detector coverage 
of LET and AMATERAS allowed us to obtain high-quality data not available in 
previous experiments \cite{MultipleMagnonModes}. The ${\vec Q}$-integrated 
intensity shown in the inset of \FigSub{fig:ExpAndSim}{h} establishes that 
the three types of bilayer are present in the approximate ratios \A:\B:\C = 
3:2:1, as detailed in Sec.~S7 of the SM \cite{SM}. This information, 
which makes completely specific the average structure reported in 
Ref.~\cite{TwoTypesOfAdjacentLayers}, is the foundation for the \textit{ABABAC}
bilayer sequence in the minimal model [\FigSub{fig:dimercell}{a}]. This 
model manifestly allows a highly accurate determination of the magnetic 
interactions and provides an excellent account of the measured spectra, 
in which neither the very weak orthorhombicity [$b/a = 1.00167(1)$ 
\cite{TwoTypesOfAdjacentLayers}\,] nor the incommensurability 
\cite{Kraemer2007, Samulon} of the crystal structure plays a role. 

The signs and sizes of the interaction parameters we deduce are fully 
consistent with the {\it ab initio} analysis \cite{Mazurenko}. Assuming 
negligible magnetostriction, these zero-field parameters have immediate 
consequences for the field-temperature phase diagram and the field-induced 
QPT. Based on the initial work of Refs.~\cite{Jaime, SebastianPRB}, 
many authors have discussed their \bacusio~data \cite{Sebastian,
MultipleMagnonModes, Kraemer2007, Kraemer2013} and models 
\cite{BatistaFischer2007, SchmalianBatista2008, RoeschVojta2007_short, 
RoeschVojta2007_long, Kamiya, LaFlorencie2009, LaFlorencie2011} by assuming 
AF intra-bilayer interactions, and hence strong inter-bilayer frustration, 
begging the question of how to understand these measurements when frustration 
is entirely absent. To address this issue in a fully quantitative manner, we 
perform state-of-the-art stochastic series expansion QMC simulations 
\cite{SSEMethod} of the six-bilayer (\textit{ABABAC}) model of Fig.~\ref{fig:dimercell} 
using the interaction parameters we determine by INS. 

\begin{figure}[t]
\begin{center}
\includegraphics[width=\columnwidth]{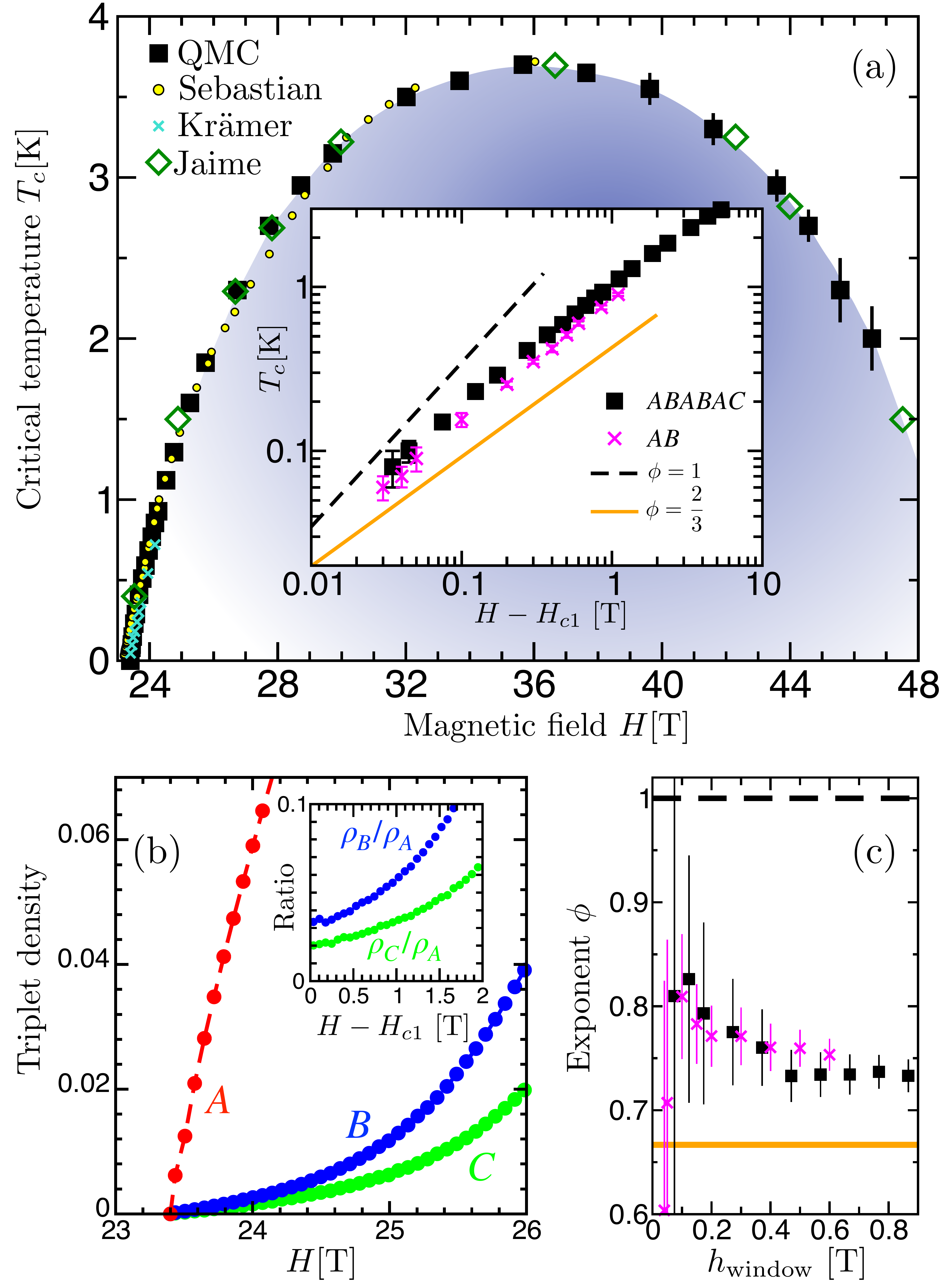}
\caption[]{QMC results for the \textit{ABABAC} model.
\FigInd{a} ($H,T$) phase diagram of the field-induced ordered regime, compared 
with experimental data from Refs.~\cite{Jaime, Sebastian, Kraemer2007}. (Inset)
Power-law scaling obtained for \textit{ABABAC} and \textit{AB} models (see text) 
and compared with the scaling forms of pure 2D ($\phi = 1$) and 3D ($\phi = 2/3$) 
criticality.
\FigInd{b} Triplet populations in the three types of bilayer, computed for 
1536 dimer units and shown as a function of field at $T = 100$ mK. (Inset)
Population ratios.
\FigInd{c} Effective exponent, $\phi$, extracted from power-law fits of the 
QMC data [inset (a)] to a window of width $h_{\rm window}$.}
\label{fig:QMC}
\end{center}
\end{figure}

We simulate an effective Hamiltonian of hard-core bosons \cite{Mila1998}, 
to reduce the computational cost, on lattices of size $3L^3$ up to $L = 22$ 
(equivalent to 63\,888 spins); details are presented in Sec.~S8 of the SM 
\cite{SM}. We calculate the temperature, $T_{c}(H)$, of the field-induced 
ordering transition from the scaled spin stiffness, $L^{(d+z-2)} \rho(L,T)$ 
\cite{Sandvik1998}, obtaining complete quantitative agreement with the phase 
boundary measured in Refs.~\cite{Jaime, Sebastian, Kraemer2007} over the entire 
range of field-induced magnetic order [\FigSub{fig:QMC}{a}]. Here we have fixed 
$g_{\parallel} = 2.435$ so that the lower critical field is $H_{c1} = 23.4$ T, 
matching the best estimate available from nuclear magnetic resonance (NMR) 
\cite{Kraemer2013}. Thus the agreement in \FigSub{fig:QMC}{a} is achieved 
with no adjustable parameters, which, even in comparison with other 
well-characterized quantum magnets \cite{Zapf}, is quite remarkable.

In \FigSub{fig:QMC}{b} we show the field-induced triplet densities in all 
three bilayers at a fixed low temperature of $T = 100$ mK. Clearly the 
\A-bilayer density, $\rho_{A}$, rises rapidly and near-linearly in $h = 
H - H_{c1}$, whereas $\rho_{B}$ and $\rho_{C}$ rise slowly with leading 
linear and quadratic components, in agreement with the NMR determination of 
$\rho_{A}$ and $\rho_{B}$ \cite{Kraemer2013}. The magnetic order parameter 
is determined by the effective triplet tunnelling between the \A bilayers, 
$t_{\rm 3D}^{\rm eff}(h)$, which is limited by the low triplet densities in the \B 
and \C bilayers (whose individual gaps have not yet closed). At low densities, 
$t_{\rm 3D}^{\rm eff}(h)$ is proportional to the ratios $r_{\rm BA} = \rho_{B}/
\rho_{A}$ and $r_{\rm CA}$ \cite{LaFlorencie2009}, and thus close to 
$H_{c1}$ takes the generic form 
\begin{equation}
t_{\rm 3D}^{\rm eff}(h) = t_{c}^{\rm eff} + a_1 h + a_2 h^2 + O(h^3).
\label{eq:t3D}
\end{equation}
$r_{BA}$ and $r_{CA}$ grow smoothly at $H > H_{c1}$ [inset, 
Fig.~\ref{fig:QMC}(b)] and, crucially, remain finite at $H = H_{c1}$, 
giving a finite $t_{c}^{\rm eff}$ (which is the case for any incomplete 
frustration \cite{RoeschVojta2007_long, Kraemer2013, LaFlorencie2011}). 
Thus as $h \to 0$, the condensed triplets have a strongly anisotropic but 
3D dispersion and the critical behavior of the phase boundary is described 
by the exponent $\phi = z/d = 2/3$ of a fully 3D QPT. 

However, the structural modulation introduces a new energy scale, 
$J_B \! - \! J_A$, and with it an energy scale $(J^{\prime\prime})^2/
(J_B \! - \! J_A)$ \cite{LaFlorencie2009}. Temperatures above this 
latter scale (approximately 0.04 K) may act to decouple the \A bilayers, 
leading to an anomalous effective exponent, $\phi(h)$, over a range of $h$. 
This is corroborated by our QMC results for $T_c(H)$ in the candidate quantum 
critical regime [inset, \FigSub{fig:QMC}{a}], both for the \textit{ABABAC} model of 
\bacusio~and for a simplified two-layer \textit{AB} model giving access to lower 
temperatures. We extract an effective $\phi$ by power-law fits in a window 
of width $h_{\rm window}$ and find [\FigSub{fig:QMC}{c}] that it does not show 
accurate 3D scaling for any realistic $h_{\rm window}$, but also never approaches 
2D scaling. Only for $h_{\rm window} < 0.08$ T do our AB-model data suggest that 
the genuine 3D critical scaling regime is entered. Above this we demonstrate 
a clear crossover into an additional regime of non-trivial effective scaling, 
arising due to bilayer modulation, in which $\phi(h)$ is determined by the 
field-evolution of $t_{\rm 3D}^{\rm eff}(h)$.

We therefore verify that the \textit{ABABAC} model with an unfrustrated inter-bilayer 
interaction is fully consistent with all of the previous, highly detailed 
experiments that have probed the properties of \bacusio~\cite{Jaime, Sparta, 
Samulon, Kraemer2007, Kraemer2013, MultipleMagnonModes, Sebastian, Zvyagin,
TwoTypesOfAdjacentLayers, Stern, SebastianPRB}. The appearance of dimensional 
reduction near the QPT \cite{Sebastian} is a consequence only of the 
inequivalent bilayer units and not of frustration \cite{RoeschVojta2007_long}. 
The unconventional physics in this previously unrecognized regime is contained 
in the effective triplet tunnelling between \A bilayers [Eq.~\eqref{eq:t3D}], 
where $t_{c}^{\rm eff} \neq 0$ ensures both a finite (if narrow) regime of 3D 
scaling (the true QPT is always 3D) and a non-universal, window-dependent 
scaling regime with $2/3 < \phi < 1$. With rapidly improving technological 
capabilities for building atomically layered magnetic materials 
\cite{Gibertini2019, Klein2019, Ubrig2019, Cao2019, Wang2019, Kim2019}, 
this type of knowledge concerning emergent behavior due to layer modulation 
will be essential to the design of their physical properties. 

In summary, we report high-resolution INS measurements over the full band 
width of the magnetic excitations in \bacusio. We have determined the 
minimal magnetic Hamiltonian required to model the spectrum and find that 
it contains three different bilayer types in the ratios 3:2:1. We verify 
that the effective intra-bilayer interaction parameters are ferromagnetic, 
which precludes any inter-bilayer frustration. We perform QMC simulations 
of the full magnetic model to demonstrate that our parameters account with 
quantitative accuracy both for the entire ($H$, $T$) phase boundary and for 
its anomalous scaling within a window of width 1 K near the QPT.  

\smallskip
\noindent
{\it Acknowledgments.} We thank S.\,Sebastian and I.\,Fisher for providing 
large, high-quality single-crystal samples and F.\,Giorgianni, M.\,Horvati\'c, 
P.\,Puphal, D.\,Sheptyakov, R.\,Stern and S.\,Ward for helpful discussions. 
N.\,L.\, and F.\,M.\, are also indebted to C.\,Berthier and S.\,Kr\"amer for an 
earlier collaboration on this material.
This work is based in part on experiments performed at the Swiss Spallation 
Neutron Source SINQ at the Paul Scherrer Institute.
Measurements on AMATERAS were performed based on the 
approved proposal No.~2015A0320.
Experiments at the ISIS Pulsed Neutron and Muon Source were supported 
by a beam time allocation from the Science and Technology Facilities Council.
We thank the Swiss National Science Foundation and the ERC Grant Hyper 
Quantum Criticality (HyperQC) for financial support. 
N.\,L.\, thanks the French National Research Agency (ANR) for support under project 
THERMOLOC ANR-16-CE30-0023-0. We acknowledge CALMIP (Grants No.~2018-P0677 
and No.~2019-P0677) and GENCI (Grant No.~2018-A0030500225) for high-performance 
computing resources.

%


\clearpage

\setcounter{equation}{0}
\renewcommand{\theequation}{S\arabic{equation}}
\setcounter{figure}{0}
\renewcommand{\thefigure}{S\arabic{figure}}
\setcounter{section}{0}
\renewcommand{\thesection}{S\arabic{section}}
\setcounter{table}{0}
\renewcommand{\thetable}{S\arabic{table}}

\onecolumngrid

\centerline{\large {\bf {Supplemental Material for ``Multiple Magnetic 
Bilayers and}}} 

\vskip1mm

\centerline{\large {\bf {Unconventional Criticality without Frustration 
in \bacusio''}}} 

\vskip4mm

\centerline{S. Allenspach, A. Biffin, U. Stuhr, G. S. Tucker, 
S. Ohira-Kawamura, M. Kofu,}

\centerline{D. J. Voneshen, M. Boehm, B. Normand, N. Laflorencie, F. Mila, 
and Ch. R\"{u}egg}

\vskip8mm

\twocolumngrid

\section{S1. Basis Transformation between Unit Cells}
\label{s1}

As stated in the main text, using the crystallographic unit cell of 
\bacusio~creates an ambiguity in the interpretation of observed triplon
dispersions. In brief, doubling of the spatial unit cell creates two 
excitation branches in the halved Brillouin zone, which have complementary 
energy minima and maxima. Here we present the transformation between the 
basis \unprimedUC of the orthorhombic crystallographic unit cell of the 
low-temperature structure, as reported in Ref.~\cite{TwoTypesOfAdjacentLayers}, 
and basis \primedUC of the unit cell of the minimal model shown in Fig.~1 of 
the main text. These differ not only in the plane of the bilayer [Fig.~1(b)] 
but also in the number of bilayers in a cell (respectively 4 and 6). 

The $c$ and $c^{\prime}$ axes are both perpendicular to the bilayers. We define 
the quantity $d_{\text{bl}}$ for the separation of neighboring bilayers, whence
\begin{equation}
c = |\vec{c}| = 4 d_{\text{bl}} \enspace \text{and} \enspace c^{\prime} = 
|\vec{c}^{\,\prime}| = 6 d_{\text{bl}}.
\label{eq:relations1}
\end{equation}
The basis vectors of the minimal and crystallographic unit cells in the plane 
of the bilayers are related by 
\begin{equation}
\vec{a} = \vec{b}^{\,\prime} + \vec{a}^{\,\prime} \enspace \text{and} \enspace 
\vec{b} = \vec{b}^{\,\prime} - \vec{a}^{\,\prime}.
\label{eq:relations2}
\end{equation}
The transformation of coordinates in reciprocal space is then given by 
\begin{equation}
Q_{h} = Q_{h}^{\prime} + Q_{k}^{\prime}, \enspace Q_{k} = Q_{k}^{\prime} - Q_{h}^{\prime}, 
\enspace Q_{l} = {\textstyle \frac{2}{3}} Q_{l}^{\,\prime} 
\label{eq:rrs1}
\end{equation}
and the inverted form 
\begin{equation}
Q_{h}^{\prime} = {\textstyle \frac12} (Q_{h} - Q_{k}), \enspace Q_{k}^{\prime} = 
{\textstyle \frac12} (Q_{h} + Q_{k}), \enspace Q_{l}^{\prime}
 = {\textstyle \frac{3}{2}} Q_{l}.
\label{eq:rrs2}
\end{equation}
We note that using a tetragonal crystallographic unit cell results in the 
same transformation. We refer to ${\vec Q}$ as the ``crystallographic basis'' 
and to ${\vec Q}^\prime$ as the ``minimal basis.'' In the halved Brillouin 
zone of each bilayer in the crystallographic basis, the second triplon mode 
is phase-shifted by half a period and is actually a ``shadow mode'' with no 
intensity, but further analysis is required to establish which mode is the 
``real'' one.

\section{S2. Single-Crystal Sample}
\label{s2}

\begin{figure}[t]
\centering
\includegraphics[width=0.9\linewidth]{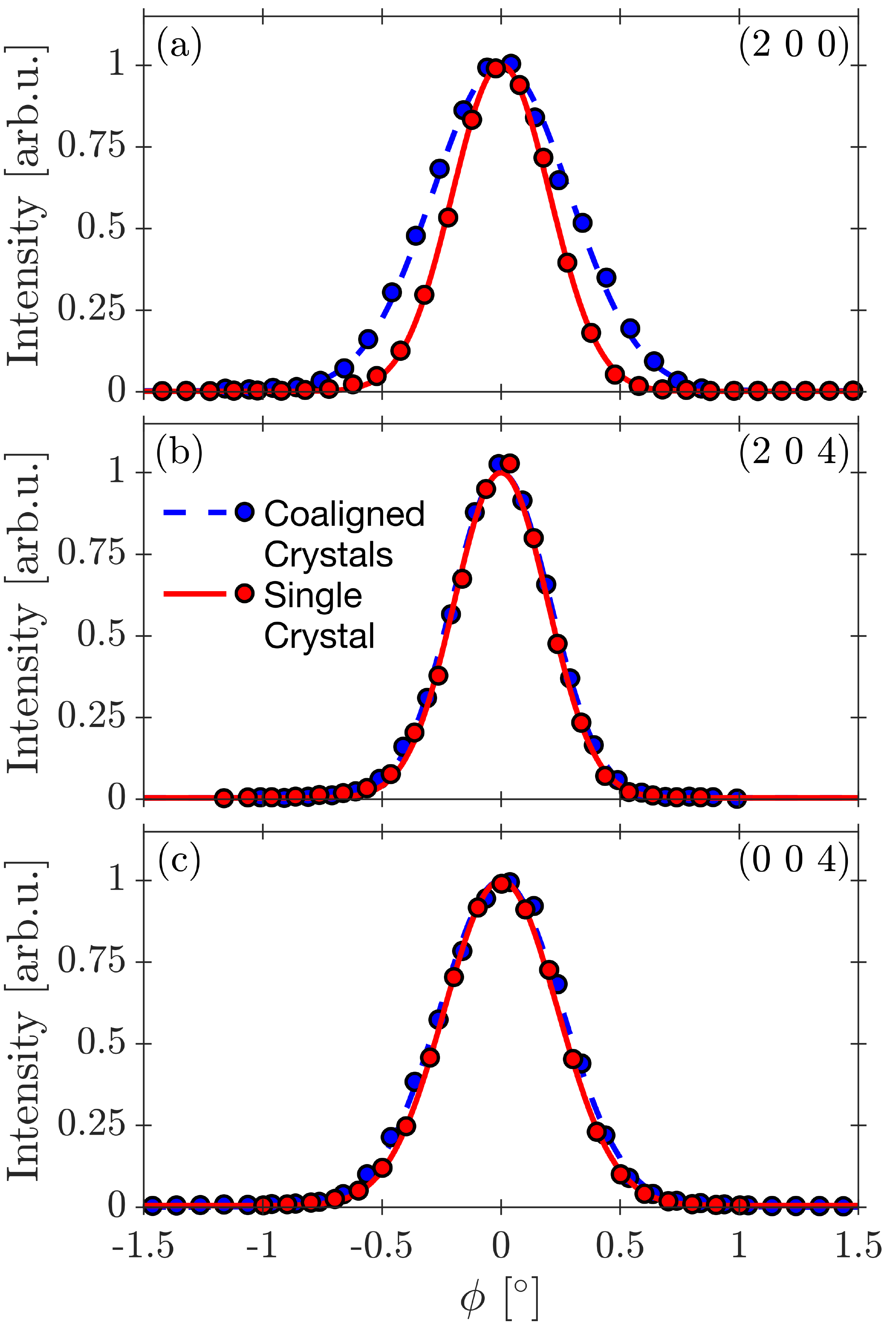}
\caption[]{Rocking curves across three Bragg peaks of the single-crystalline 
\bacusio~sample (solid red lines), which are (a) (2 0 0), (b) (2 0 4), and 
(c) (0 0 4) in the crystallographic basis. Shown for comparison are rocking 
curves obtained with the same instrument settings for the coaligned sample 
of five small single crystals used in Ref.~\cite{MultipleMagnonModes}. The 
intensities of all peaks have been scaled to unity.}
\label{fcr}
\end{figure}

For the experimental measurements on all spectrometers we used the same 
flux-grown single crystal, which was oblate in form and had a mass of 1.01 g. 
The growth method and basic characterization information are presented in 
Ref.~\cite{SebastianPRB}. For the purposes of a scattering experiment, we 
measured the mosaic spread of the crystal at room temperature on the two-axis 
diffractometer MORPHEUS at the SINQ neutron source; the same procedure was 
employed to characterize the sample used in Ref.~\cite{MultipleMagnonModes}, 
which consisted of five smaller coaligned crystals with a total mass of 
1.25 g. Figure~\ref{fcr} shows rocking curves across two Bragg peaks for 
both samples. These scans were performed with the same resolution settings 
in both cases, in order to be able to compare the peak shapes, but we 
caution that these were not precisely the same settings used in Fig.~2(a) 
of Ref.~\cite{MultipleMagnonModes}. While Fig.~\ref{fcr} indicates that the 
structural quality of the single-crystal sample is slightly superior to that 
of the coaligned one, additional advantages for INS purposes include its 
compact nature and no aluminium holder, all of which resulted in significant 
improvements in energy and wave-vector resolution.

\section{S3. Data Pre-Processing}
\label{s3}

Raw TOF data from LET were pre-processed using the software Mantid 
\cite{Mantid} and data from AMATERAS using Utsusemi \cite{Utsusemi}.
On LET, the raw data were normalized to a vanadium standard and an empty-can 
measurement was subtracted. The software package Horace \cite{Horace} was 
then used to create ``cuts'' in wave vector, $\vec{Q}$, and energy transfer, 
$\omega$, from the four-dimensional dataset. All data gathered on both ToF 
and triple-axis instruments were multiplied by the factor $k_{i}/k_{f}$ to 
obtain the intensities, $I({\vec Q},\omega)$, presented in 
Figs.~2(a)-2(d) and 3 of the main text. 

To analyze the ToF data, we prepared cuts as a function of $\omega$ 
for sets of $\vec{Q}$ points along selected high-symmetry directions.
These cuts displayed two or three independent excitation peaks, depending on 
whether modes \B and \C were distinguishable at the relevant $\vec{Q}$, which  
were fitted with the corresponding number of Gaussian functions plus a 
background contribution. In \FigOneDimCuts of the main text, the maxima of 
the Gaussians are taken to define the mode energies and their weights as 
the mode intensities.

\section{S4. Modeling the Spectrum}
\label{s4}

It is clear both from the structure of \bacusio~\cite{TwoTypesOfAdjacentLayers}
and from earlier experimental results \cite{MultipleMagnonModes} that there 
exist at least two different dimer types in the unit cell, or dimer sublattices 
within the system. Dimers in separate sublattices have different intradimer 
interaction parameters, and presumably also different interdimer (but 
intra-sublattice) ones. To calculate the magnetic excitation spectrum under 
these circumstances we consider an arbitrary dimerized $S = 1/2$ spin system 
with only Heisenberg interactions. 

\begin{figure}[t]
\centering
\includegraphics[width=0.75\linewidth]{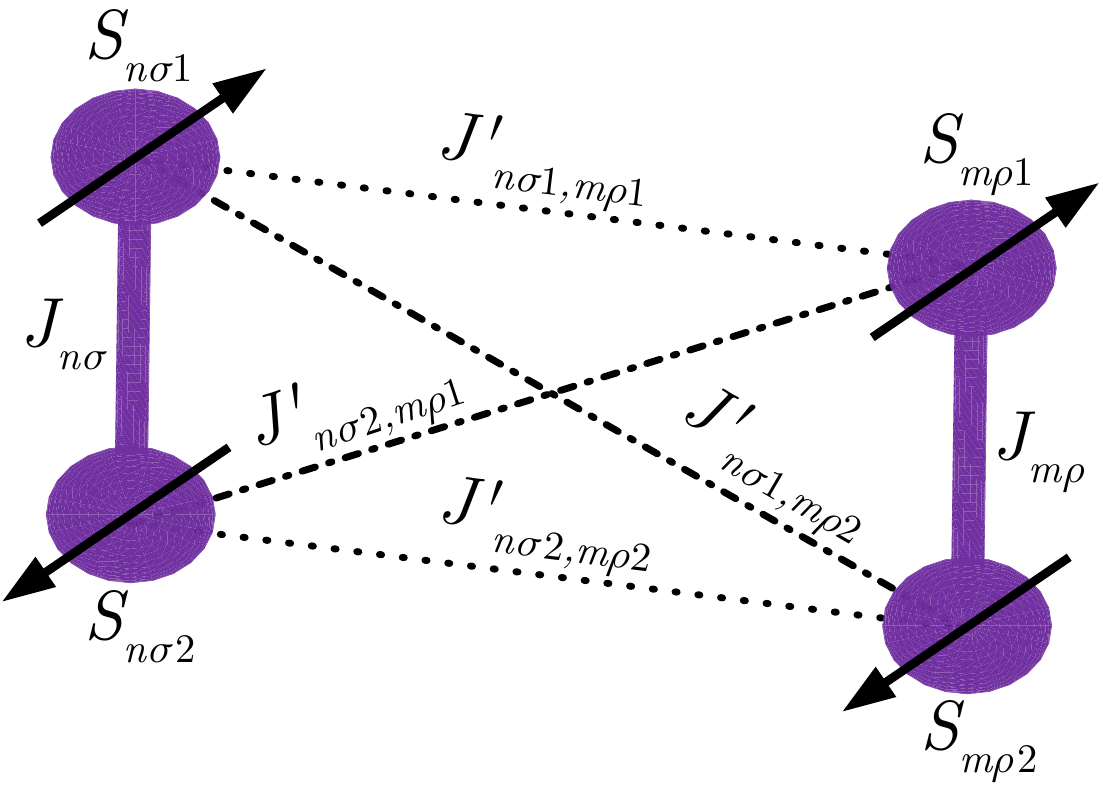}
\caption[]{Schematic representation of the interaction pathways between 
two pairs of dimerized $S = 1/2$ ions. Arrows represent instantaneous 
relative spin orientations for all-AF Heisenberg interactions obeying 
$J_{n \sigma}, J_{m \rho} > J^\prime_{n \sigma 1, m \rho 2}, J^\prime_{n \sigma 2, m \rho 1}
 > J^\prime_{n \sigma 1, m \rho 1}, J^\prime_{n \sigma 2, m \rho 2}$. }
\label{fii}
\end{figure}

The magnetic Hamiltonian can be expressed as
\begin{eqnarray}
\label{eq:GeneralHamiltonianRPA}
\mathcal{\hat{H}} & = & \mathcal{\hat{H}}_{\text{intra}} + 
\mathcal{\hat{H}}_{\text{inter}} \\ & = & \sum_{n,\sigma} J_{n\sigma} \, 
\mathbf{\hat{S}}_{n \sigma 1} \! \cdot \! \mathbf{\hat{S}}_{n \sigma 2} \; + \!\! 
\sum_{nm, \sigma \rho, pr}^\prime {\textstyle \frac12} J_{n \sigma p, m \rho r}^\prime 
\, \mathbf{\hat{S}}_{n \sigma p} \! \cdot \! \mathbf{\hat{S}}_{m \rho r}, \nonumber
\end{eqnarray}
in which $\mathbf{\hat{S}}_{n \sigma p}$ is the operator for the ionic spin on 
a single site, $m$ and $n$ label separate unit cells, $\sigma$ and $\rho$ 
label the dimer sublattice within a unit cell, $p,r\in\{1,2\}$ label the 
two ions making up each dimer, and the second sum excludes the term $n 
\sigma = m \rho$ (which is in fact the first sum). The eigenbasis of 
$\mathcal{\hat{H}}_{\text{intra}}$ consists only of the singlet and triplet
eigenstates of a single dimer, whose total occupations must sum to unity.
Because there is no external magnetic field, the three triplet modes are 
degenerate at all times. The starting condition that the dimer bonds, 
$J_{n\sigma}$, are AF and the strongest interactions in the system ensures 
that the problem is reduced to the dynamics of a very low density of triplets 
moving in a singlet background. As represented in Fig.~\ref{fii}, 
the effective interdimer hopping of these triplon quasiparticles is given in 
terms of the ionic interactions of Eq.~(\ref{eq:GeneralHamiltonianRPA}) by 
\begin{align}
\begin{split}
J^\prime_{n \sigma, m \rho} = & \;\; {\textstyle \frac12} \big( J^\prime_{n \sigma 1, 
m \rho 1} - J^\prime_{n \sigma 1, m \rho 2} \\ & \;\;\;\;\;\; - J^\prime_{n \sigma 2, 
m \rho 1} + J^\prime_{n \sigma 2, m \rho 2} \big).
\label{general_dimerexchange}
\end{split}
\end{align}
Clearly in the case of the intra-bilayer interactions labeled $J^\prime_\sigma$
($\sigma =$ \A, \B, \C) in Fig.~1(a) of the main text, the effectively FM sign 
of these parameters results from AF ionic interaction terms satisfying 
$J^\prime_{n \sigma 1, m \rho 2} = J^\prime_{n \sigma 2, m \rho 1} > J^\prime_{n \sigma 1, 
m \rho 1} = J^\prime_{n \sigma 2, m \rho 2}$ \cite{Mazurenko}. We comment that 
the inter-bilayer term, $J^{\prime\prime}$, is treated as a single, ionic 
interaction (Fig.~1(a) of the main text), and hence the sign of this 
parameter is in fact the consequence of a genuinely FM superexchange process.

To discuss the triplet dynamics, we generalized the Green-function approach 
of Refs.~\cite{Leuenberger1983, Leuenberger1984}, which for any number, $N$, 
of sublattices provides $2N$ independent coupled equations for diagonal and 
off-diagonal processes in the triplet occupation basis, but does not mix 
triplet modes at different wave vectors, ${\vec K}$~\cite{Haley}. Assuming 
at zero temperature that the singlet and triplet occupation probabilities 
on every dimer obey $\langle \hat{n}^{\sigma}_{s} \rangle + 3\langle 
\hat{n}^{\sigma}_{t} \rangle = 1$ is an approximation entirely equivalent to 
using the bond-operator approach \cite{rnr} in the Holstein-Primakoff 
approximation \cite{rmnrs}. The poles of the Green function yield the 
triplon dispersion relations for each $\vec{K}$. 

We restrict our considerations to spin dimer systems, as described by the 
Hamiltonian of Eq.~(\ref{eq:GeneralHamiltonianRPA}), but with only one dimer 
type in each bilayer, because structures involving different dimer types in 
the same bilayer would exhibit spectra completely different from those observed 
in experiment. In the remainder of this section we use ${\vec K}$ to denote the 
wave vector of a general bilayer dimer system and connect our expressions to 
\bacusio~at the end. For a system with only one dimer type ($\sigma
 = A$), one obtains 
\begin{equation}
\omega_{A}(\vec{K}) = \sqrt{J_A^{2} + J_{A} J_{AA} (\vec{K})},
\label{eq:onedimerdispersion}
\end{equation}
where $J_{AA} (\vec{K})$ is obtained from the terms appearing in 
Eq.~(\ref{general_dimerexchange}) under Fourier transformation of 
$\mathcal{\hat{H}}$. With two dimer bilayers ($\sigma = A$ and 
$\rho = B$), the dispersion relations are 
\begin{equation}
\omega_{\pm} (\vec{K}) = \sqrt{ {\textstyle \frac12} [ \omega^{2}_{A}
(\vec{K}) + \omega^{2}_{B} (\vec{K}) \pm \mathcal{J} (\vec{K}) ]}
\label{eq:twodimerdispersions}
\end{equation}
with
\begin{equation}
\mathcal{J} (\vec{K}) = \sqrt{ [\omega^{2}_{A} (\vec{K}) - \omega^{2}_{B} 
(\vec{K})]^2 + 4J_{A} J_{B} J_{AB}^2 (\vec{K})},
\label{eq:twodimerdispersions_additional}
\end{equation}
in which we use $\omega_{A} (\vec{K})$ and $\omega_{B} (\vec{K})$ from 
Eq.~(\ref{eq:onedimerdispersion}) for brevity. For more than two dimer 
sublattices, the polynomial expression has a degree higher than four and 
it is more convenient to find its roots numerically.

The explicit forms of the functions $J_{\sigma\rho} (\vec{K})$ in these 
expressions are written most easily by separating intra- and inter-bilayer 
interaction terms. The intra-bilayer terms in a two-bilayer model ($\sigma 
\in \{\A,\B\}$) are given by
\begin{equation}
J_{\sigma\sigma} (\vec{K}) = 2J^{\prime}_{\sigma} [\cos ( 2 \pi K_h ) + \cos ( 2 \pi 
K_k ) ],
\label{eintra}
\end{equation}
while the inter-bilayer term for adjacent bilayers of types $\sigma = \A$ and 
$\rho = \B$ is 
\begin{equation}
J_{AB} (\vec{K}) = - 4J^{\prime\prime} \cos (\pi K_h) \cos (\pi K_k) \cos (\pi K_l).
\label{einter}
\end{equation}

By constructing the generalized susceptibility tensor 
\cite{NeutronScatteringInCondensedMatterPhysics} from the triplon Green 
functions \cite{Leuenberger1984}, one obtains the magnetic dynamical 
structure factor at zero temperature, which for a system with one dimer 
type is given by 
\begin{equation}
\mathcal{S}^{\alpha\beta} (\vec{K},\omega) = N \delta^{\alpha\beta} \big[ f_A 
(\vec{K}) / \omega_{A}(\vec{K}) \big] \, \delta (\omega - \omega_{A} (\vec{K})),
\label{eq:single_dimer_structurefactor}
\end{equation}
where $f_{A} (\vec{K}) = J_{A} [1 - \cos (\vec{K} \! \cdot \! \vec{R}_{A} )]$, 
with $\vec{R}_{A}$ the vector connecting the ions of dimer $A$. For a system 
with two dimer bilayers,
\begin{eqnarray}
\label{efi}
\mathcal{S}^{\alpha\beta} (\vec{K},\omega) = & & {\textstyle \frac{1}{4}} 
N \delta^{\alpha\beta} \big[ f_{A} (\vec{K}) + f_{B}(\vec{K}) \big] 
\\ \times \bigg[ \big( 1 \! + \! \Xi(\vec{K}) \big) && \frac{\delta \big( 
\omega \! - \! \omega_{+} (\vec{K}) \big)}{\omega_{+} (\vec{K})} + \big( 1 \! - 
\! \Xi (\vec{K}) \big) \frac{ \delta \big( \omega \! - \! \omega_{-} (\vec{K}) 
\big)}{\omega_{-} (\vec{K})} \bigg], \nonumber
\end{eqnarray}
where
\begin{equation}
\Xi (\vec{K}) = \frac{\big[ f_{A} (\vec{K}) \! - \! f_{B} (\vec{K}) \big] \big[ 
\omega^{2}_{A} (\vec{K}) \! - \! \omega^{2}_{B} (\vec{K}) \big] \! + \! f_{AB} 
(\vec{K})} {\big[ f_{A} (\vec{K}) \! + \! f_{B} (\vec{K}) \big] \mathcal{J} 
(\vec{K})}
\end{equation}
with 
\begin{align}
\begin{split}
f_{AB} (\vec{K}) = \; & 8 J_{A} J_{B} \sin \big( {\textstyle \frac12} 
\vec{K} \! \cdot \! \vec{R}_{A} \big) \sin \big( {\textstyle \frac12} \vec{K} 
\! \cdot \! \vec{R}_{B} \big) J_{AB} ( \vec{K}).
\end{split}
\end{align}
In \refequ{efi}, $\omega_\pm (\vec{K})$ is given in 
\refequ{eq:twodimerdispersions} and we show only the part of the dynamical 
structure factor resulting from excitation processes with positive energy 
transfer [$\omega (\vec{K}) > 0$]. 

For a system with more than two bilayer types, which includes the case 
of \bacusio~(Fig.~1 of the main text), the analytical expression is too 
cumbersome to be useful and we employ a numerical procedure to construct 
$\mathcal{S}^{\alpha\beta}(\vec{K},\omega)$ for every point $\vec{K}$. We 
performed these calculations using the minimal basis, $\vec{Q}^{\prime}$ in 
Sec.~S1, but for historical and experimental consistency we show all our 
results in the crystallographic basis, $\vec{Q}$. When the illustrative 
calculations above are extended to the minimal \textit{ABABAC} model 
for \bacusio, 
the dispersion relations shown in Figs.~2(a)-2(b), 2(e)-2(f), and 3(a) of the 
main text are dominated by intra-bilayer terms of the form shown in 
\refequ{eintra} for $\sigma \in \{\A,\B,\C\,\}$. For the inter-bilayer terms, 
one may use \refequ{einter} with 
\begin{equation}
K_{h} = Q_{h}^{\prime}, \enspace K_{k} = Q_{k}^{\prime}, 
\enspace K_{l} =  {\textstyle \frac13} Q_{l}^{\prime}
\end{equation}
to see using \refequ{eq:rrs2} that in the crystallographic basis 
\begin{equation}
\label{einter_crystallographic}
J_{AB} (\vec{Q}) \propto [ \cos ( \pi Q_h ) + \cos ( \pi Q_k ) ], 
\end{equation}
and hence that all modes are completely flat along \Qlldir, as shown in 
Figs.~2(d) and 2(h) of the main text.

\begin{figure}[t]
\centering
\includegraphics[width=0.8\linewidth]{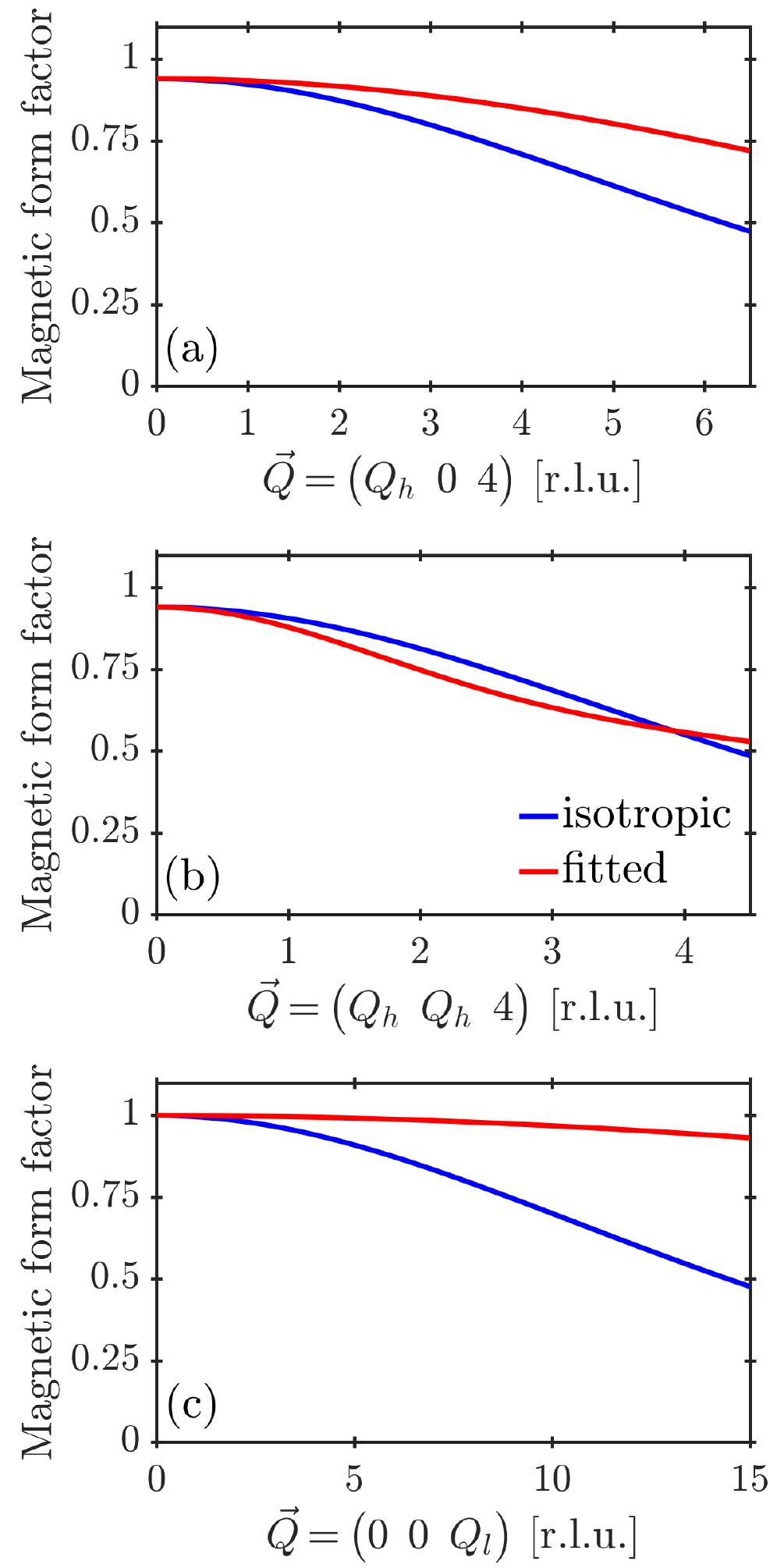}
\caption[]{
Magnetic form factor shown for the high-symmetry $\vec{Q}$ directions (a) 
\Qhdir, (b) \QhQhdir, and (c) \Qldir in the crystallographic basis. Results 
for an isotropic spin density on the Cu$^{2+}$ ions (solid blue lines) are 
compared with results obtained from optimal fits of the full scattered 
intensity in the measured data range (solid red lines), as described in the 
text. The magnetic form factor is normalized to 1 at ${\vec Q}$ = (0 0 0).}
\label{fig:mffc}
\end{figure}

\section{S5. Magnetic Form Factor}
\label{s5}

The form factor describing magnetic scattering from ions with part-filled 
3$d$ and 4$d$ electron shells, which are assumed to produce a spatially 
isotropic spin density, can be approximated in an intermediate 
$|\vec{Q}|$ range by 
\cite{ITC_VolumeC}
\begin{equation}
F_{\text{iso}}(s) \approx \langle j_{0}(s) \rangle \approx \tilde{A} e^{-\tilde{a}s^{2}} 
+ \tilde{B}e^{-\tilde{b}s^{2}} + \tilde{C}e^{-\tilde{c}s^{2}} + \tilde{D},
\label{isomagneticformfactor}
\end{equation}
where the variable $s$ is defined as
\begin{equation}
s = \frac{\sin(\theta)}{\lambda} = \frac{|\vec{Q}|}{4\pi}.
\label{definition_of_s}
\end{equation}
The parameters $\{\tilde{A}, \tilde{a}, \tilde{B}, \tilde{b}, \tilde{C}, \tilde{c}, 
\tilde{D}\}$ are ion-specific and are tabulated in Ref.~\cite{ITC_VolumeC},
where they are expressed without the tilde symbols (added here to avoid 
confusion with the bilayer labels and unit-cell parameters of our work).

The process of modelling the measured cross section for (coherent) inelastic 
magnetic scattering at low temperatures requires multiplying the magnetic 
dynamical structure factor by the square of the magnetic form factor. 
In our calculations of this cross section for the high-symmetry ${\vec Q}$ 
directions measured in experiment, using the isotropic magnetic form factor 
for Cu$^{2+}$ gives intensity functions that fall differently as functions of 
$|\vec{Q}|$ when compared to the data, as Fig.~\ref{fig:mffc} makes clear. 
Thus we conclude from our results that the spin density of the Cu$^{2+}$ ions 
in \bacusio~must indeed be anisotropic \cite{Mazurenko,Zvyagin}; this could 
be a consequence of hybridization effects that concentrate charge in the 
bilayer planes, analogous to those observed in the low-dimensional Cu$^{2+}$ 
compounds \lacuo~\cite{La2CuO4_Paper1, La2CuO4_Paper2} and 
\srcuo~\cite{SrCuO2_Paper1, SrCuO2_Paper2}.

The consequence of this anisotropy is that the form factor falls more slowly 
with $|\vec{Q}|$ for the \Qhdir and \Qldir directions than an isotropic 
function would dictate [Figs.~\ref{fig:mffc}(a) and \ref{fig:mffc}(c)], 
whereas such behavior is not found along \QhQhdir [Fig.~\ref{fig:mffc}(b)]. 
To account for this effect, we have fitted our data using three separate 
functions with the form of \refequ{isomagneticformfactor} for each of the 
directions \Qhdir, \QhQhdir, and \Qldir while keeping fixed the value of 
the dynamical structure factor in our calculation.

\begin{figure*}[p] 
\begin{center}
\includegraphics[width=\textwidth]{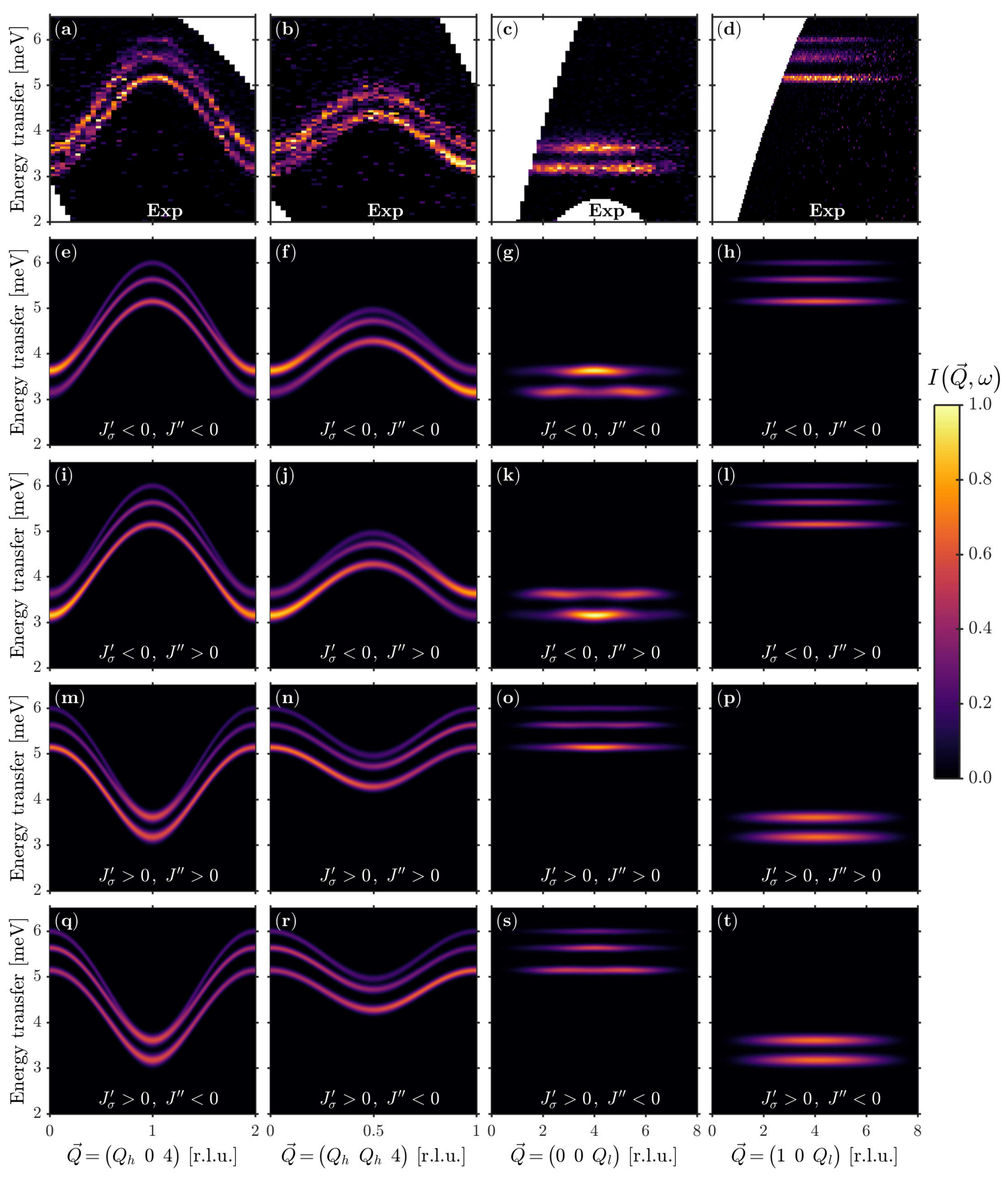}
\caption[]{
\FigIndRange{a}{d} Experimental spectrum for four ${\vec Q}$ directions, 
reproduced for reference from Fig.~2 of the main text. \FigIndRange{e}{t} 
Corresponding spectrum calculated by using the minimal model of Fig.~1(a) of 
the main text and with the magnitudes of the Heisenberg interaction parameters 
deduced for \bacusio, but taking different signs for the three intra-bilayer 
($J^{\prime}_{\sigma}$, $\sigma \in \{\A,\B,\C\,\}$) and for the inter-bilayer 
($J^{\prime\prime}$) interaction parameters. We show only cases where all three 
intra-bilayer interaction parameters have the same sign. \FigIndRange{e}{h} FM  
$J^{\prime}_\sigma$ and $J^{\prime\prime}$, as in Figs.~2(e)-2(h) of the main 
text. \FigIndRange{i}{l} FM $J^{\prime}_\sigma$ and AF $J^{\prime\prime}$. 
\FigIndRange{m}{p} AF $J^{\prime}_\sigma$ and $J^{\prime\prime}$. 
\FigIndRange{q}{t} AF $J^{\prime}_\sigma$ and FM $J^{\prime\prime}$.}
\label{fig:differentsigncases}
\end{center}
\end{figure*}

\section{S6. Different signs of the interaction parameters}
\label{s6}

Changing the signs of the intra- and inter-bilayer interaction parameters 
has no effect on the density of triplon states. Thus the majority of 
experiments performed on \bacusio, including specific heat, magnetization, 
magnetic torque, magnetocalorimetry, and NMR, depend strongly on the 
magnitudes of these parameters but are very insensitive to their signs, 
which is one of the reasons why the sign of the intra-bilayer term has 
remained ambiguous for so long. Because the signs of these parameters 
were the crucial point of ambiguity in interpreting the earlier INS 
experiments, which we resolve in the present work, and because they 
are crucial to the frustration mechanism of dimensional reduction, in 
\Fig{fig:differentsigncases} we present calculations of the INS spectra 
that would be measured in \bacusio~under each possible scenario for AF 
or FM interactions. We consider only the physical modes of the minimal 
unit cell and show for reference the experimental spectra [Figs.~2(a)-2(d) 
of the main text]. 

Figures \ref{fig:differentsigncases}(e)-\ref{fig:differentsigncases}(f), 
\ref{fig:differentsigncases}(i)-\ref{fig:differentsigncases}(j), 
\ref{fig:differentsigncases}(m)-\ref{fig:differentsigncases}(n), 
and \ref{fig:differentsigncases}(q)-\ref{fig:differentsigncases}(r) 
illustrate the clear inversion of the in-plane mode dispersions caused 
by changing the intra-bilayer interactions from FM to AF. Specifically, the 
minima (the ``bilayer gaps'') occurring in the crystallographic basis at even 
($Q_{h}$ 0) and integer ($Q_{h}$ $Q_{h}$) for FM $J^{\prime}$ become maxima for AF 
$J^{\prime}$, and conversely at odd ($Q_{h}$ 0) and half-integer ($Q_{h}$ $Q_{h}$). 

Figures \ref{fig:differentsigncases}(g), \ref{fig:differentsigncases}(k), 
\ref{fig:differentsigncases}(o), and \ref{fig:differentsigncases}(s) illustrate 
the behavior of the two mode intensities in the almost non-dispersive 
\Qldir spectrum, where the double-peak structure is obtained for mode \A 
when $J^{\prime\prime}$ is FM, whereas for AF $J^{\prime\prime}$ it appears for 
modes \B and \C; in each case the other modes have a single intensity peak. 
Figures \ref{fig:differentsigncases}(h), \ref{fig:differentsigncases}(l), 
\ref{fig:differentsigncases}(p), and \ref{fig:differentsigncases}(t) show 
that the completely non-dispersive mode energy along \Qlldir is independent 
of the sign of $J^{\prime\prime}$ [\refequ{einter}], and in fact occurs in all 
circumstances when tetragonal symmetry of the inter-bilayer interactions is 
maintained. The same statement may be made for the effectively $Q_l$-invariant 
mode intensity, which contains only the dimer form factor.

Before leaving this discussion, we comment briefly on the comparison 
between our optimal interaction parameters and those of previous neutron 
scattering studies. In the early work of Ref.~\cite{Sasago}, the data 
were fitted to a dimer model containing one dimer type and including only 
a next-nearest-neighbor intra-bilayer interaction, $J_{2}^{\prime}$, while 
the nearest-neighbor intra-bilayer term, our $J_{\sigma}'$, was set to zero.
Although the intradimer interaction obtained by these authors, $J = 4.41(2)$ 
meV, is close to the average of the $J_{\sigma}$ parameters we determine, the 
FM $J_{2}^{\prime} = - 0.19(3)$ gives a dispersion far from the one we measure. 
With only these two parameters, the intra-bilayer spin structure consists 
of two disconnected but interpenetrating lattices and thus is not fully 
determined. 

In the later work by some of the present authors~\cite{MultipleMagnonModes}, 
as noted in the main text and in Sec.~S1, the analysis was affected by the 
sign ambiguity, caused by using the crystallographic unit cell, and by the 
historical assumption of AF intra-bilayer interactions. The intradimer and 
intra-bilayer interaction parameters of this study are close in magnitude to 
our present results, but with larger error bars. In addition to refining the 
accuracy and signs of these terms, here we obtain entirely new information 
in the size and sign of the inter-bilayer interaction, $J^{\prime\prime}$. 
Because the size of $J^{\prime\prime}$ we deduce exceeds the upper limit thought 
to be consistent with the results of Ref.~\cite{MultipleMagnonModes}, our 
minimal model does not require any modulation of the inter-bilayer interaction,
of the type proposed in Ref.~\cite{RoeschVojta2007_long}, to explain any of 
the available \bacusio~data. We stress again that the sign of $J^{\prime\prime}$ 
is not the crucial ingredient delivering inter-bilayer frustration, as this 
is present for either sign of $J^{\prime\prime}$ if the $J_{\sigma}^{\prime}$ 
parameters are AF, but for neither when they are FM. 

\section{S7. Relative fractions of bilayer types}
\label{s7}

For a more general expression of the case \Qlldir of the previous subsection, 
in any structurally dimerized material with more than one dimer type, one may 
define $\vec{Q}^{\star}$ as the subset of $\vec{Q}$ space where the Fourier 
transform of the interactions between dimers on different sublattices is 
identically zero. For $\vec{Q}$ points in $\vec{Q}^{\star}$, the dimer 
sublattices are effectively decoupled and one may approximate the inelastic 
magnetic cross section, using the dynamical structure factor for each separate 
dimer sublattice [Eq.~(\ref{eq:single_dimer_structurefactor})], by
\begin{equation}
\frac{d^{2}\sigma}{d\Omega d\omega} (\vec{Q},\omega)\Big|_{\vec{Q} \, \in \, 
\vec{Q}^{\star}} = \mathcal{N}(\vec{Q},\omega) \sum_{\rho} \frac{N_{\rho}J_{\rho}} 
{\omega_{\rho}(\vec{Q})} \, \delta \big( \omega - \omega_{\rho} (\vec{Q}) \big),
\label{esbt}
\end{equation}
where all terms independent of the dimer sublattice, including the Debye-Waller
factor, the magnetic form factor, and $k_f/k_i$, are contained in the prefactor 
$\mathcal{N}(\vec{Q},\omega)$. Convolving the $\delta$ functions with the 
instrumental resolution function of the spectrometer yields finite-width 
line shapes that are usually approximated as Gaussians.

\begin{table}[t]
\centering
\begin{tabular}{c | c | c }
$N^{\text{rel}}_{A} \enspace[\%] \;$ & $\; N^{\text{rel}}_{B} \enspace[\%] \;$ & 
$\; N^{\text{rel}}_{C} \enspace[\%] \;$ \\
\hline
$51(5)$ & $32(4)$ & $17(2)$\\
\end{tabular}
\caption{Relative intensities of excitations corresponding to bilayer types 
\A, \B, and \C.}
\label{tab:RelativefractionsTable}
\end{table}

\bacusio~presents an especially straightforward case where the separate dimer 
sublattices ($\rho$) are the \A, \B, and \C bilayers. From Eqs.~(\ref{einter}) 
and (\ref{eq:rrs2}) one may read that $\vec{Q}^{\star}$ includes all points with 
$Q_h^\prime = Q_k^\prime = Q_l^\prime/3 = \pm (2n + 1)/2$, where $n$ is an integer. 
More specifically, from \refequ{einter_crystallographic} it is clear that 
\Qlldir is contained within $\vec{Q}^{\star}$ for all $Q_{l}$. \Qlldir has the 
additional advantage that the \A, \B, and \C modes are maximally separated in 
energy (Fig.~2 of the main text). Because the triplon modes along \Qlldir 
are completely flat, the statistical accuracy can be increased by integrating 
over a very wide range in $Q_{l}$. Using our data from AMATERAS, we extracted 
the function $I^{\star} (\omega)$ by integrating over [0.95, 1.05] in $Q_{h}$, 
[$-0.05$, 0.05] in $Q_{k}$, and [4, 8] in $Q_{l}$; in the inset of Fig.~2(h) of 
the main text, we chose to label the $Q_l$ coordinate by $\bar{Q}_l$ to reflect 
this extraordinarily wide integration window. As this inset makes clear, 
$I^{\star} (\omega)$ displays three narrow and well-separated peaks, which we 
fit using three Gaussians (plus a background term). By integrating each 
Gaussian over its full energy, we obtain one intensity parameter, $I_\rho$, 
for each bilayer.

By using Eq.~(\ref{esbt}) we then determine the relative fractions of the 
three different dimer types using
\begin{equation}
N^{\text{rel}}_{\rho} = \frac{N_{\rho}}{\sum_{\eta} N_{\eta}} = \frac{I_{\rho} \,
\omega_{\rho} / J_{\rho}}{\sum_{\eta} I_{\eta} \, \omega_{\eta} / J_{\eta}}, 
\label{eq:relative_fractions}
\end{equation}
where the sublattice energy scale, $J_\rho$, is the intradimer interaction 
and $\omega_{\rho}$ is the energy of each non-dispersive band in Figs.~2(d) 
and 2(h) of the main text. The results for $N^{\text{rel}}_{\rho}$ are displayed 
in Table \ref{tab:RelativefractionsTable} and verify at the 5\% accuracy 
level that the bilayer ratio is indeed 3:2:1. We stress that this analysis 
is completely independent of the bilayer stacking order and the actual 
value of the inter-bilayer interaction parameter, $J^{\prime\prime}$.

\begin{figure}[t]
\centering
\includegraphics[width=0.96\columnwidth,clip]{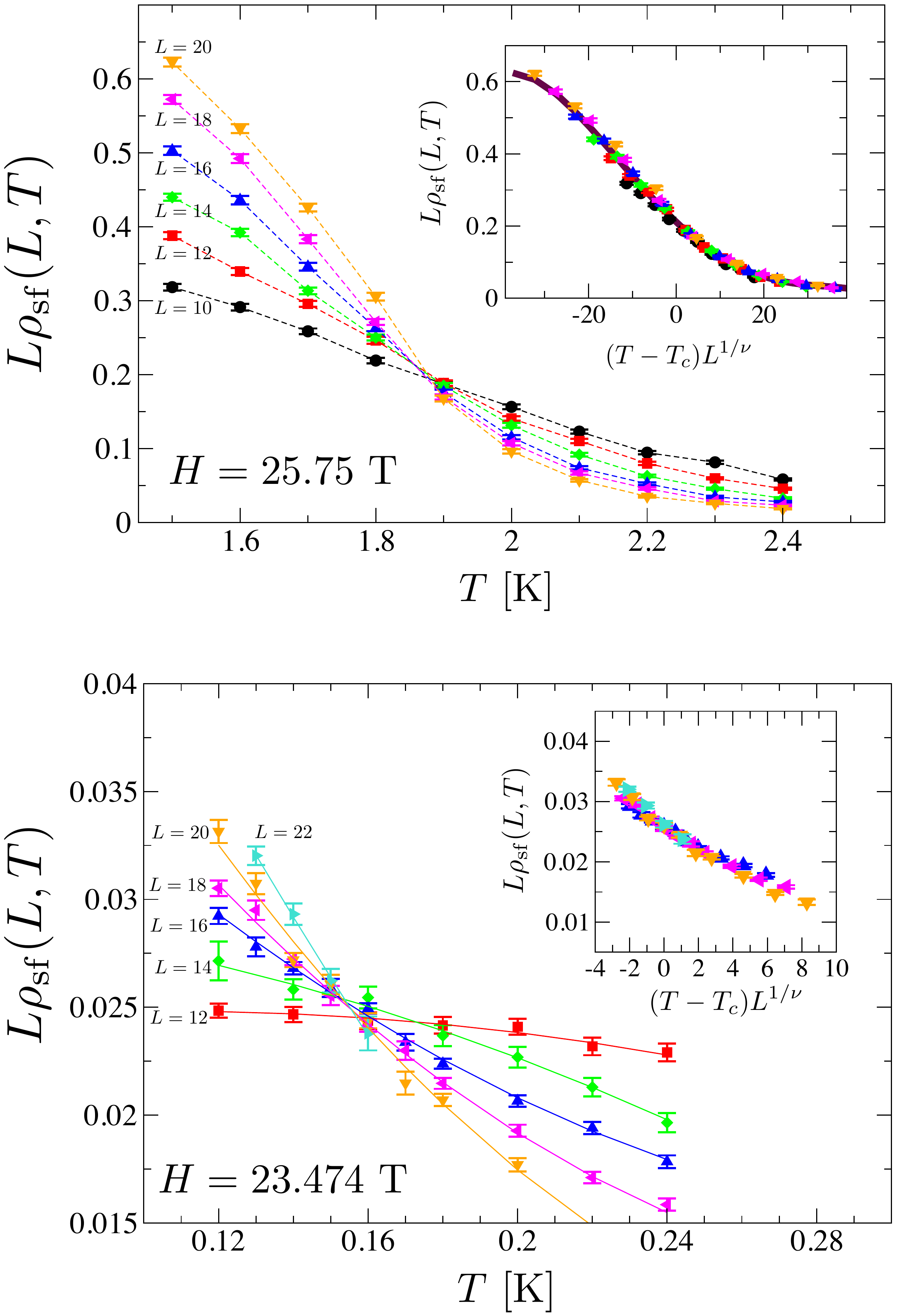}
\caption{Scaled spin stiffness, $L \rho_{\rm sf} (L,T)$, shown for all values 
of $L$ in our simulations to illustrate the unique crossing point for two 
values of the magnetic field. The extracted value of $T_c$ allows the scaling 
collapse shown in the insets. As $H$ approaches $H_{c1}$ (lower panel), 
$\rho_{\rm sf}$ becomes very small and a large relative error becomes 
inevitable.}
\label{fig:cross}
\end{figure}

\section{S8. Quantum Monte Carlo}
\label{s8}

We have performed large-scale QMC simulations using the stochastic series 
expansion (SSE) algorithm~\cite{SSEMethod} on 3D cubic lattices of sizes 
$L$$\times$$L\times$$L_\perp$, with $L_\perp = 3L$, for $L = 8$, 10, 12, \ldots, 
22. The critical temperature, $T_c$, is extracted for each applied field, $H$, 
by using the finite-size analysis of the superfluid density, $\rho_{\rm sf}$, of 
the hard-core bosons representing the triplons. In the magnetic context, the 
superfluid density is equivalent to a spin stiffness. Because $L^{d + z - 2} 
\rho_{\rm sf}(L,T)$ has scaling dimension zero, $d = 3$ here and $z = 0$ for a 
finite-temperature transition, all curves $L \rho_{\rm sf} (L,T)$ for different 
$L$ cross at the same point, $T = T_c$ \cite{Sandvik1998}, as illustrated for 
two values of $H$ in Fig.~\ref{fig:cross}. With $T_c$ fixed in this way, one 
may extend similar arguments to discuss the thermal scaling, which is governed 
by the finite-size form~\cite{Sandvik1998}
\begin{equation}
\rho_{\rm sf} (L,T) = L^{2-d} {\cal{F}} [L^{1/\nu} (T - T_c)].
\end{equation}
As the insets of Fig.~\ref{fig:cross} make clear, good data collapse is 
achieved using a value $\nu = 0.66$ for the critical exponent of the 
correlation length, in excellent agreement with the best available estimate 
for this universality class (3D XY) of $\nu = 0.672$ \cite{Campostrini2000}.

As shown in Fig.~4(a) of the main text, our QMC simulations provide an 
excellent quantitative account of the entire $(H,T)$ phase boundary of 
\bacusio. In the challenging regime near the field-induced QPT at $H_{c1}$, 
the physics of the system is determined by the mobility of the triplets 
condensing in the \A bilayers at fields where the \B and \C bilayers, 
whose associated dispersion gaps have not yet closed (Fig.~2 of the main 
text), contain only low, ``proximity-induced'' triplet densities (Fig.~4(b) 
of the main text). From Eq.~(1) of the main text, perfectly frustrated 
inter-bilayer coupling~\cite{LaFlorencie2009} would in fact allow $t_{c}^{\rm 
eff} = 0$, no field-linear component in $\rho_{B}$ or $\rho_{C}$, and 
possible $\phi = 1$ scaling over some range of $h_{\rm window}$. Nevertheless, 
the true QPT is always a 3D phenomenon, because at $T$ = 0 K the system does 
develop spatial coherence in all three dimensions on some low energy scale. 

Away from perfect frustration, and in particular for \bacusio, $t_{c}^{\rm eff} 
\neq 0$ ensures both 3D and anomalous ($h_{\rm window}$-dependent) scaling over 
regimes of finite width. We comment that the windowing procedure is extremely 
sensitive to the chosen value of $H_{c1}$; this is not an issue in our present 
calculations, because $H_{c1}$ is fixed by the interactions determined from 
INS, but when $H_{c1}$ is also an unknown to be fitted simultaneously 
\cite{Sebastian}, changes of 0.01 T can cause changes of order 0.1 in $\phi$. 

\end{document}